# Isabelle/HOL as a Meta-Language for Teaching Logic

Asta Halkjær From      Jørgen Villadsen

DTU Compute - Department of Applied Mathematics and Computer Science - Technical University of Denmark

Patrick Blackburn

Section of Philosophy and Science Studies, IKH, Roskilde University, Denmark

Proof assistants are important tools for teaching logic. We support this claim by discussing three formalizations in Isabelle/HOL used in a recent course on automated reasoning. The first is a formalization of System W (a system of classical propositional logic with only two primitive symbols), the second is the Natural Deduction Assistant (NaDeA), and the third is a one-sided sequent calculus that uses our Sequent Calculus Verifier (SeCaV). We describe each formalization in turn, concentrating on how we used them in our teaching, and commenting on features that are interesting or useful from a logic education perspective. In the conclusion, we reflect on the lessons learned and where they might lead us next.

## 1 Introduction

Today's logical landscape contains classical and non-classical, propositional and higher-order, extensional and intensional, constructive and infinitary, and many other systems, both implemented and abstract. Making sense of all this requires a grasp of such concepts as syntax versus semantics, differing proof styles and their tradeoffs, translations and embeddings, and interactions between different levels of language and proof. We claim that proof assistants are an important tool for teaching logic, as they make such architectural issues explicit right from the start, and do so in a way that makes them accessible even to relatively inexperienced students.

We illustrate this by discussing three formalizations in Isabelle/HOL. The first is a formalization of System W, a system for propositional logic due to Mordchaj Wajsberg [6, footnote 259] that dates back to 1937, the second is the Natural Deduction Assistant (NaDeA), and the third is a one-sided sequent calculus that uses our Sequent Calculus Verifier (SeCaV). These were used in a recent course on automated reasoning which used Isabelle/HOL [16, 17] as a key teaching resource.[1] The use of Isabelle/HOL forced students to grapple with the modern logical architecture — but it also provided them with a smooth proof environment which enabled even our (relatively inexperienced) students to explore themes such as soundness and completeness successfully.

We will discuss each of these formalizations in turn, starting with System W, a propositional system, and then moving on to NaDeA and SeCaV, both of which handle classical first-order logic. Our discussion will highlight aspects of these formalizations that we think are pedagogically important. Some of these are relatively concrete, such as the concise notation we use in our teaching for writing sequent calculus derivations; others are more abstract, like the use of "proof search search" where the system helps write the proofs. In the conclusion we draw the threads together, reflecting on pedagogical lessons learned and where they may lead us next.

---

[1] The course was taught by the first two authors at the Technical University of Denmark during the Spring 2020 semester (the last part of the course was taught via Zoom due to the COVID-19 shutdown); 27 students signed up for the exam in June.

Our work is part of the IsaFoL (Isabelle Formalization of Logic) project that aims at developing formalizations in Isabelle/HOL of logics, proof systems, and automatic/interactive provers [5]. The formalization of NaDeA and SeCaV can be obtained from the NaDeA online web application as described later (in total 6498 lines in Isabelle/HOL). System W is available on GitHub: https://github.com/logic-tools/axiom.

One final remark: using Isabelle/HOL rather than say Coq [3, 18] or Agda [8, 24] is already a pedagogical choice, but one we made for purely pragmatic reasons: it is the proof assistant we know best.

## 2 System W in Isabelle/HOL

As Isabelle is a *generic* proof assistant, it offers two main ways of specifying a logic. First, we can specify it as an axiomatization in Isabelle's logical framework; this allows us to work on proofs within our logic, but denies us the ability to talk about which proofs are possible. We considered it pedagogically preferable to take the second option: embedding System W in the higher-order logic of Isabelle/HOL. Doing so leads to a second choice point: should we embed shallowly, treating its syntax as a subset of the metalogical syntax, or deeply, where we define its syntax as objects in the metalogic with an explicit semantics function? Again, the latter approach seemed pedagogically preferable since we wanted to teach as much metatheory as possible, and we can only teach structural induction on the syntax (an important part of any logic education) if it has a concrete representation in the metalogic.[2]

---

**datatype** *form = Falsity* (⟨⊥⟩) | *Pro nat* | *Imp form form* (**infix** ⟨→⟩ 0)

---

**inductive** *Axiomatics* (⟨⊢⟩) **where**
 ⟨⊢ q⟩ **if** ⟨⊢ p⟩ **and** ⟨⊢ (p → q)⟩ |
 ⟨⊢ (p → (q → p))⟩ |
 ⟨⊢ ((p → q) → ((q → r) → (p → r)))⟩ |
 ⟨⊢ (((p → q) → p) → p)⟩ |
 ⟨⊢ (⊥ → p)⟩

**abbreviation** *Truth* (⟨⊤⟩) **where** ⟨⊤ ≡ (⊥ → ⊥)⟩

**theorem** ⟨⊢ ⊤⟩ **using** *Axiomatics.intros*(5) .

**primrec** *semantics* (**infix** ⟨⊨⟩ 0) **where**
 ⟨(I ⊨ ⊥) = False⟩ |
 ⟨(I ⊨ (Pro n)) = I n⟩ |
 ⟨(I ⊨ (p → q)) = (if I ⊨ p then I ⊨ q else True)⟩

**theorem** ⟨I ⊨ p⟩ **if** ⟨⊢ p⟩ **using** *that* **by** *induct auto*

**definition** ⟨*valid p* ≡ ∀I. (I ⊨ p)⟩

**theorem** ⟨*valid p* = ⊢ p⟩ **oops**

---

Here is a small snapshot of the material. The **datatype** command recursively defines the syntax of formulas. The proof system is then defined as a predicate over formulas using the **inductive** command: the first line gives the modus ponens rule, the remaining lines the axiom schemas. The semantics is given as a primitive recursive predicate on the syntax with a definition for each constructor. Incidentally: Isabelle gives a warning if any case is left out. An **abbreviation** is expanded automatically while a **definition** introduces a new name that can be unfolded at will. Note that validity is defined, and completeness stated, with the **oops** command indicating that details need to be filled in.

System W is a very simple propositional system — but already some useful points are emerging. First: the formalization *forces* the student to take all these components and their interaction seriously. In particular, right from the start the student is face-to-face with the syntax/semantics distinction and is naturally led to the concepts of soundness and completeness. Second: it *supports* them in doing this, in ways ranging from gentle reminders about missing cases, to carrying out "proof search search". For example, the **sledgehammer** command can search for (and find) proof search methods that can find proofs of $\vdash \top$, and $\vdash p \to p$. And in the completeness proof, it can establish all cases of Hintikka model existence once we have chosen to do the proof by induction. Third: it even makes it *fun*. As Dominic Mulligan, Tobias Nipkow and Vladimir Voevodsky have all remarked, using a proof assistant is a bit like playing a very complex video game [14, 15]. The code is there, it's simple, and the curious student can play with it, for example by experimenting with other choices of connectives and axioms.

---

[2]In a sense we find ourselves at the meta-meta-meta level. Our object of study, System W, sits at the bottom, specified in higher-order logic (meta). This is again specified in Isabelle's logical framework (meta-meta) and finally we describe it in natural language (meta-meta-meta).

## 3  The Natural Deduction Assistant (NaDeA)

There are several parts to our Natural Deduction Assistant (NaDeA). First of all, we have a formalization in Isabelle of the syntax and semantics of classical first-order logic and a natural deduction proof system. Our syntax has falsity as a primitive and defines negation in terms of this and implication. We represent constants as functions taking no arguments and variables as natural numbers referring to them using de Bruijn indices. The proof rules are defined inductively and we have formalized proofs of soundness and completeness. Since the proof system is formalized in Isabelle we can be extremely precise about the side conditions of rules and the substitution procedure [22, 23].

The NaDeA online web application is built on top of this formalization:

```
https://nadea.compute.dtu.dk/
```

The formalization document `Natural_Deduction_Assistant.thy` can be obtained by clicking on the verification button in the top right corner when the help window has been cancelled. Note that the verification button itself shows the number of ¤ symbols in the current proof state (initially 1) and if this becomes 0 then the proof is finished and a proof in Isabelle/HOL is generated. The formalization document `Natural_Deduction_Assistant.thy` is found in the so-called "Base theory" tab.

The online web application is written in the TypeScript programming language and allows the user to input a formula and to prove it using the proof system. It has several features [23]:

- The user is only presented with rules applicable to solving the chosen subgoal.
- The application automatically keeps track of assumptions and appeals to them.
- Side conditions of quantifier rules are checked, i.e. that Skolem constants are new.
- The user can undo and redo to any previous state.
- Proofs, whether finished or in-progress, can be exported to a textual format that can be loaded again later or on another computer. This preserves all the steps taken in the proof.
- The user can switch back and forth between the standard notation and the abstract syntax used in the formalization.
- A version of the formalization can be viewed inside the application with comments alongside the proof system definitions and the soundness proof.

Another aspect of the web application is a system dubbed ProofJudge [21]. This allows the teacher of a logic course to pose formulas as exercises or assignments that the students can access, try to prove, and hand in. The teacher and teaching assistants can then see how many people have solved a given task, see individual solutions, and so on. ProofJudge integrates with the assistant so that you can see how many steps a student used to prove a given formula or how many subgoals are still left. The system also allows students to save their work online and return to it later [21].

While ProofJudge is designed to facilitate the human grading of submissions, we also run an automatic tableau prover in the background as the user works on their proof. This prover checks every current subgoal and if it seems likely that the subgoal is unprovable, it unobtrusively gives a warning by making the corresponding line number orange [22]. We have verified the soundness of the prover's kernel in Isabelle [11, 12] and used the code export facilities to generate SML code for the full prover. From this we generate the JavaScript that runs on the page [22].

Going from NaDeA to Isabelle, we have a feature that allows users to export their finished online proofs to a corresponding proof in the Isabelle formalization. There, each rule application including side conditions is checked by the proof assistant and our formalized soundness proof then guarantees the validity of the formula. The possibility of exporting proofs in this way mitigates the fact that the web application is not formally verified: if you are in doubt of the validity of your proof you can export it to Isabelle and have it checked there [22].

We have evaluated our use of NaDeA in the classroom and note that a problem for small proofs is that students can potentially find them by just clicking blindly, though this is not a problem for more complicated examples [20]. One feature appreciated by several students is the ability to access any previous state through undoing and redoing, a feature that is not present in most applications where taking an action after undoing makes it impossible to go back. A number of examples and hints are available in the system to get started [20].

## 4  The Sequent Calculus Verifier (SeCaV)

A recent spinoff of the NaDeA project is the Sequent Calculus Verifier (SeCaV). Work on SeCaV was started by the first two authors in November 2019 and since then it has played a role in our teaching, some of which has previously been described [9]. SeCaV uses the same syntax as NaDeA, but the proof system is a one-sided sequent calculus. We will now describe this calculus, and the kind of problems we have set for our students using it.

We represent sequents as lists of formulas. The calculus is one-sided and as such has rules not just for each connective and quantifier, but also for the negation of each connective and quantifier. The proof system is given in Figure 1. It is important to note that *Neg* is not primitive but defined as *Neg p* ≡ *Imp p Falsity*. This is why we can make do without any rule for double negation; it is covered by the implication rules.

The rules are classified using Smullyan's well-known uniform notation [19]. Propositional rules that do not branch are called $\alpha$-rules and start with *Al*. Propositional rules that branch start with *Be* for $\beta$. The *Ga*-rules, for $\gamma$, operate on quantified formulas that can be built from arbitrary instances, i.e. existential statements and negated universals. To derive a formula with the *De*-rules, for $\delta$, the constant used in the derived instance must be new. Note that all the rules work on the first formula in the sequent. This makes the rule easier to state and easier for the simplifier to work with, which in turn makes formalizing derivations smoother. Also note that the *Basic* axiom allows us to derive any sequent whose head occurs negated somewhere in the tail.

Two additional rules are worth commenting on. First, the derived *Neg* rule allows us to remove double negations:

**theorem** *Neg*: ⟨⊩ *Neg* (*Neg p*) # *z*⟩ **if** ⟨⊩ *p* # *z*⟩
 *(proof omitted)*

**inductive** *sequent-calculus* (⟨⊩ - ⟩ *0*) **where**
　*Basic*: ⟨⊩ *p* # *z*⟩ **if** ⟨*member* (*Neg p*) *z*⟩ |
　*AlDis*: ⟨⊩ *Dis p q* # *z*⟩ **if** ⟨⊩ *p* # *q* # *z*⟩ |
　*AlImp*: ⟨⊩ *Imp p q* # *z*⟩ **if** ⟨⊩ *Neg p* # *q* # *z*⟩ |
　*AlCon*: ⟨⊩ *Neg* (*Con p q*) # *z*⟩ **if** ⟨⊩ *Neg p* # *Neg q* # *z*⟩ |
　*BeCon*: ⟨⊩ *Con p q* # *z*⟩ **if** ⟨⊩ *p* # *z*⟩ **and** ⟨⊩ *q* # *z*⟩ |
　*BeImp*: ⟨⊩ *Neg* (*Imp p q*) # *z*⟩ **if** ⟨⊩ *p* # *z*⟩ **and** ⟨⊩ *Neg q* # *z*⟩ |
　*BeDis*: ⟨⊩ *Neg* (*Dis p q*) # *z*⟩ **if** ⟨⊩ *Neg p* # *z*⟩ **and** ⟨⊩ *Neg q* # *z*⟩ |
　*GaExi*: ⟨⊩ *Exi p* # *z*⟩ **if** ⟨⊩ *sub 0 t p* # *z*⟩ |
　*GaUni*: ⟨⊩ *Neg* (*Uni p*) # *z*⟩ **if** ⟨⊩ *Neg* (*sub 0 t p*) # *z*⟩ |
　*DeUni*: ⟨⊩ *Uni p* # *z*⟩ **if** ⟨⊩ *sub 0* (*Fun c* []) *p* # *z*⟩ **and** ⟨*news c* (*p* # *z*)⟩ |
　*DeExi*: ⟨⊩ *Neg* (*Exi p*) # *z*⟩ **if** ⟨⊩ *Neg* (*sub 0* (*Fun c* []) *p*) # *z*⟩ **and** ⟨*news c* (*p* # *z*)⟩ |
　*Extra*: ⟨⊩ *z*⟩ **if** ⟨⊩ *p* # *z*⟩ **and** ⟨*member p z*⟩

Figure 1: The sequent calculus.

Second, although we have the *Extra* rule which allows us to drop a head that already exists elsewhere, in practice it is more useful to use the admissible *Ext* rule that rearranges, contracts or adds formulas:

**theorem** *Ext*: ⟨⊩ *y*⟩ **if** ⟨⊩ *z*⟩ **and** ⟨*ext y z*⟩
*(proof omitted)*

Here, *ext y z* expresses that *y* is an extension of *z* in that it contains all the formulas that *z* does and possibly more:

**primrec** *ext* **where**
⟨*ext y* [] = *True*⟩ |
⟨*ext y* (*p* # *z*) = (**if** *member p y* **then** *ext y z* **else** *False*)⟩

The function *member* is defined straightforwardly:

**primrec** *member* :: ⟨*fm* ⇒ *fm list* ⇒ *bool*⟩ **where**
⟨*member p* [] = *False*⟩ |
⟨*member p* (*q* # *z*) = (**if** *p* = *q* **then** *True* **else** *member p z*)⟩

The *ext* relation is equivalent to the subset relation:

**lemma** *member* [*simp*]: ⟨*member p z* ⟷ *p* ∈ *set z*⟩
　**by** (*induct z*) *simp-all*

**lemma** *ext* [*simp*]: ⟨*ext y z* ⟷ *set z* ⊆ *set y*⟩
　**by** (*induct z*) *simp-all*

The proof of completeness for this system is derived from a completeness proof for a tableau system whose rules are the dual of the sequent calculus. These tableau rules correspond closely to the consistency property conditions used in a formalization by Berghofer [2]. This means that we can apply his result to show completeness for the tableau system. We then translate any closing tableau into a sequent calculus derivation (for the negated formula) and obtain completeness of the sequent calculus in this way [9]. This method showcases an important feature of proof assistants: the ability to build on top of other people's work with complete confidence that you apply their results correctly. We do not need to formalize an entire completeness proof for SeCaV; instead we employ strategic translations between proof systems to make an existing result applicable. Translating proofs between proof systems to transfer results from one to the other is an important technique; it is something students should be exposed to early, and Isabelle/HOL provides a good environment for teaching it.

## 5  Integrating SeCaV and NaDeA

For the Spring 2020 course, we integrated SeCaV and NaDeA more closely: they were in the same Isabelle theory file, and the two systems used exactly the same datatype to represent the syntax of formulas. This meant that students could experiment with the two systems within exactly the same environment instead of, for instance, having to learn two different ways of inputting the syntax of formulas or applying rules of inference. This allowed us to illustrate some connections between SeCaV and NaDeA and in particular, how assumptions in natural deduction translate to our one-sided sequents.

Figure 2 shows the NaDeA proof system where *OK p z* means that *p* can be derived from assumptions *z*. Unlike Figure 1 we use the standard Isabelle meta-implication rather than the **if** construct to specify the rules. This exposes students to different types of Isabelle notation.

**inductive** *OK* :: ⟨*fm* ⇒ *fm list* ⇒ *bool*⟩ **where**
 *Assume*: ⟨*member p z* ⟹ *OK p z*⟩ |
 *Boole*: ⟨*OK Falsity* (*Imp p Falsity* # *z*) ⟹ *OK p z*⟩ |
 *Imp-E*: ⟨*OK* (*Imp p q*) *z* ⟹ *OK p z* ⟹ *OK q z*⟩ |
 *Imp-I*: ⟨*OK q* (*p* # *z*) ⟹ *OK* (*Imp p q*) *z*⟩ |
 *Dis-E*: ⟨*OK* (*Dis p q*) *z* ⟹ *OK r* (*p* # *z*) ⟹ *OK r* (*q* # *z*) ⟹ *OK r z*⟩ |
 *Dis-I1*: ⟨*OK p z* ⟹ *OK* (*Dis p q*) *z*⟩ |
 *Dis-I2*: ⟨*OK q z* ⟹ *OK* (*Dis p q*) *z*⟩ |
 *Con-E1*: ⟨*OK* (*Con p q*) *z* ⟹ *OK p z*⟩ |
 *Con-E2*: ⟨*OK* (*Con p q*) *z* ⟹ *OK q z*⟩ |
 *Con-I*: ⟨*OK p z* ⟹ *OK q z* ⟹ *OK* (*Con p q*) *z*⟩ |
 *Exi-E*: ⟨*OK* (*Exi p*) *z* ⟹ *OK q* (*sub 0* (*Fun c* []) *p* # *z*) ⟹ *news c* (*p* # *q* # *z*) ⟹ *OK q z*⟩ |
 *Exi-I*: ⟨*OK* (*sub 0 t p*) *z* ⟹ *OK* (*Exi p*) *z*⟩ |
 *Uni-E*: ⟨*OK* (*Uni p*) *z* ⟹ *OK* (*sub 0 t p*) *z*⟩ |
 *Uni-I*: ⟨*OK* (*sub 0* (*Fun c* []) *p*) *z* ⟹ *news c* (*p* # *z*) ⟹ *OK* (*Uni p*) *z*⟩

Figure 2: The natural deduction proof system.

The following theorem relates NaDeA and the sequent calculus. In NaDeA, we prove formulas under some assumptions: the formula is only required to hold when all the assumptions are discharged. It is an implication on the meta-level. In the corresponding sequent all the assumptions become negated: the sequent is provable if you can either falsify an assumption (prove its negation) or show the conclusion. As intended, when all the assumptions are true, the conclusion must be too.

**theorem** *OK-sequent-calculus*: ⟨*OK p z* ⟷ (⊩ *p* # *map Neg z*)⟩
 *(proof omitted)*

As a corollary we can consider the case of no assumptions:

**corollary** ⟨*OK p* [] ⟷ (⊩ [*p*])⟩
 **unfolding** *OK-sequent-calculus* **by** *simp*

Unlike the relationship between the sequent calculus and the interim tableau system that we use for completeness, we do not show this correspondence via a translation between the systems. Instead, we use the independent soundness and completeness of both systems to reason about the relationship between the provable judgements in NaDeA and SeCaV, respectively. The intention is to strengthen the students' understanding of the two types of judgements: NaDeA works on an implication from conjoined assumptions to a conclusion, while SeCaV works on one-sided sequents.

## 6   Derivations in the Sequent Calculus

Let us look at some derivations in the calculus. Consider the following formula:

**proposition** ⟨ *p a* ⟶ *p a* ⟩ **by** *metis*

Converting to our abstract syntax we can start the derivation like so:

**lemma** ⟨⊩
[
  *Imp* (*Pre* ″*p*″ [*Fun* ″*a*″ []]) (*Pre* ″*p*″ [*Fun* ″*a*″ []])
]
⟩
**proof** −

We can neatly derive this formula by combining the **from**, **with**, **have** and **if** commands in Isabelle with the abbreviation for the goal, *?thesis*. Thus, we can apply rules that break down the formula until we reach a sequent covered by *Basic*. First the *AlphaImp* rule gives the new subgoal (we have introduced synonyms like *AlphaImp* for *AlImp* and use the synonyms in the derivations):

  **from** *AlphaImp* **have** *?thesis* **if** ⟨⊩
  [
    *Neg* (*Pre* ″*p*″ [*Fun* ″*a*″ []]),
    *Pre* ″*p*″ [*Fun* ″*a*″ []]
  ]
  ⟩
    **using** *that* **by** *simp*

Next we apply the *Ext* rule to swap the order of the two resulting formulas:

  **with** *Ext* **have** *?thesis* **if** ⟨⊩
  [
    *Pre* ″*p*″ [*Fun* ″*a*″ []],
    *Neg* (*Pre* ″*p*″ [*Fun* ″*a*″ []])
  ]
  ⟩
    **using** *that* **by** *simp*

And by doing so we have arrived at a *Basic* sequent, completing the derivation:

  **with** *Basic* **show** *?thesis*
    **by** *simp*
**qed**

An interesting feature of our proof system is that the γ-rules are "destroyed" when we instantiate them in the sub-derivation. For some proofs, however, you need several instances of the same formula. The partial derivation in Figure 3 is such an example. In this case we can start off by using the *Ext* rule to duplicate the formula, and by doing so, effectively instantiate it twice. Or if we view the derivation as going from the axioms towards the final formula, we end the derivation by using *Ext* to contract the two copies.

Another thing worth pointing out about the example in Figure 3 is the application of the *GammaExi* rule. The simplifier, as invoked by *simp*, is powerful enough to handle every rule application in our derivations except for some rule applications that involve substitution. In those cases we need to explicitly instantiate the rule with the term used in the substitution by using the **where** attribute.

The full version of Figure 3 is given in the Appendix alongside two other examples. These also showcase the use of **and** to manage branching derivations.

*A. H. From, J. Villadsen & P. Blackburn* 25```
lemma ‹⊩
  [
    Exi (Uni (Dis (Pre ''p'' [Var 0]) (Neg (Pre ''p'' [Var 1]))))
  ]
 ›
proof −
  from Ext have ?thesis if ‹⊩
    [
      Exi (Uni (Dis (Pre ''p'' [Var 0]) (Neg (Pre ''p'' [Var 1])))),
      Exi (Uni (Dis (Pre ''p'' [Var 0]) (Neg (Pre ''p'' [Var 1]))))
    ]
   ›
    using that by simp
  with GammaExi[where t=‹Fun ''a'' []›] have ?thesis if ‹⊩
    [
      Uni (Dis (Pre ''p'' [Var 0]) (Neg (Pre ''p'' [Fun ''a'' []]))),
      Exi (Uni (Dis (Pre ''p'' [Var 0]) (Neg (Pre ''p'' [Var 1]))))
    ]
   ›
    using that by simp
  ⋮
```

Figure 3: A $\gamma$-formula can be instantiated twice by duplicating it.

## 7 Teaching Sequent Calculus

We tested the alpha version of SeCaV in November 2019 as previously described [9]. We had 52 students but the SeCaV exercises were optional. NaDeA was not used.

We then tested the beta version with the NaDeA and SeCaV integration in the Spring 2020 course with 27 students. Here we had a number of mandatory assignments with exercises in both NaDeA and SeCaV. We'll start with a really challenging example that we tried out with our students. This example is discussed on page 128 of the *Handbook of Tableau Methods* [7]:

> *If every person that is not rich has a rich father, then some rich person must have a rich grandfather.*

Formalization with *r* (*rich*) and *f* (*father*):

$$\forall x(\neg r(x) \to r(f(x))) \to \exists x(r(x) \wedge r(f(f(x))))$$

Only one student managed to complete the proof for the challenge. Afterwards we reduced the proof to 19 steps in the Sequent Calculus Verifier (SeCaV) and these steps take up 466 lines in Isabelle/HOL in the format shown in the previous section and in the Appendix. We have a solution in the Natural Deduction Assistant (NaDeA) with 42 steps and 162 clicks in the web application [20].

Besides this difficult challenge, we also asked the students to work on 24 formulas and their proofs. For the take-home exam the students were asked, among other things, to prove the following 9 formulas using the formalization in Isabelle/HOL:

**proposition** ‹ $(p \longrightarrow q) \longrightarrow p \longrightarrow q$ › **by** *metis*

**proposition** ‹ $p \longrightarrow (p \longrightarrow q) \longrightarrow q$ › **by** *metis*



**proposition** ⟨$p \wedge (p \longrightarrow q) \longrightarrow q$⟩ **by** *metis*

**proposition** ⟨$p\ a \wedge (p\ a \longrightarrow (\forall x.\ p\ x)) \longrightarrow (\forall x.\ p\ x)$⟩ **by** *metis*

**proposition** ⟨$(\forall x.\ p\ x) \longrightarrow p\ a$⟩ **by** *metis*

**proposition** ⟨$p \longrightarrow q \longrightarrow p$⟩ **by** *metis*

**proposition** ⟨$(p \longrightarrow q \longrightarrow r) \longrightarrow (p \longrightarrow q) \longrightarrow p \longrightarrow r$⟩ **by** *metis*

**proposition** ⟨$(\forall x.\ p\ x) \longrightarrow (\exists x.\ p\ x)$⟩ **by** *metis*

**proposition** ⟨$p \vee (p \longrightarrow q)$⟩ **by** *metis*

24 students handed in their solutions and in general the solutions were excellent (some of the proofs were unnecessarily long). The final grades for the automated reasoning course were as follows: 10 As, 10 Bs, 4 Cs and 2 Fs (in the ECTS grading scale; one student was released from the exam). The course evaluation is available online: https://kurser.dtu.dk/course/02256/info

In August 2020 we released SeCaV 1.0 to be used without NaDeA (like the alpha version [9] but unlike the beta version described in the present paper): https://github.com/logic-tools/secav

SeCaV 1.0 has a full separation using multiple Isabelle theory files of the simpler soundness proof and the much more advanced completeness proof. SeCaV 1.0 also has a diverse collection of sample sequent calculus proofs in propositional logic as well as in first-order logic.

## 8 Discussion

Jasmin Christian Blanchette recently remarked that the automated reasoning community has largely stood on the sidelines of developments in proof assistants, preferring to *'reflexively turn to "pen and paper" — by which we usually mean LATEX to define our logics, specify our proof systems, and establish their soundness and completeness'* [5]. In a similar vein, we urge logic teachers to become early adopters of proof assistants.

A traditional approach to teaching logic is to start by giving the student experience in working with a proof calculus (often a natural deduction or tableau system) and then to show them how contemporary logical architecture fits together. Unfortunately, many students never reach the goal of "seeing" logical architecture, as these metatheoretic aspects are often only taught in more advanced courses that are only taken by (and only suitable for) students with considerable mathematical maturity. Moreover, although many beginners are now given their first steps in natural deduction or tableau systems using some visual web interface — which is certainly a step forward from simply doing "pen and paper" proofs — such systems are often rigid. They may achieve limited goals well (for example, training large groups of undergraduate students in propositional logic) and they often make life easier for overworked instructors (they may provide automated checking and grading of assignments) but it is not clear that they open the doors very far towards a deeper understanding of what logic is about.

Proof assistants put metatheory front and center — and they also enable relative novices to explore it. Teaching logic using a proof assistant like Isabelle/HOL makes fundamental architectural concepts vivid. The distinction between syntax and semantics, object and metalogics, and so on, are foregrounded right from the start. Moreover, they are presented in a technological setting that shows that they are ideas to be *explored*. Abstract questions (*How can languages and their semantics be altered?*) turn into concrete investigations (*Let's see what happens if we have three truth values instead of two!*).

Crucially, all this is done in a setting that does *not* presuppose mathematical maturity — though it is clearly a setting in which mathematical maturity can be *developed*. Isabelle/HOL is good in this regard.



As we have already noted in our System W discussion, we can use the **oops** command to discontinue the current proof but we can also use the **sorry** command instead. As it says in the manual:

> **sorry** is a fake proof pretending to solve the pending claim without further ado ... The most important application of **sorry** is to support experimentation and top-down proof development [25].

Such tools are useful to beginners: they provide support in exploring the big picture. Of course, using **sorry** clearly runs the risk of basing a proof on an approach that isn't ultimately going to work out. But this is *precisely* the sort of judgement that students have to learn to make; it is an important component of "mathematical maturity".

Moreover, Isabelle/HOL offers some good tools for filling in the missing details in the "proof search search" process. Most Isabelle proofs are not written using the primitive axioms but by invoking the proper proof search methods. Here too the system can help: for example, the **sledgehammer** tool will search for a method that finds a proof for the current goal, maybe with the help of local and library lemmas. It can sometimes help with tricky Hilbert-style proof details (*can I really derive $p \to p$ from these axioms?*) but it can also help with completeness proofs (the student who realizes that a Hintikka model existence lemma will need to be proved by induction, even if she is not sure how this should be done, knows enough to successfully invoke **sledgehammer**). On the other hand, sometimes we are in the dark: *is my idea true or not?* Here the **nitpick** and **quickcheck** commands that search for a counterexample to a proposed lemma can help (these are run automatically). Again — such tools help beginners get to grips with the architecture of proof, learning how to break them down, how to put them together, and above all, learning how to explore. Performing "proof search search" using such tools as **oops**, **sorry**, **sledgehammer, nitpick** and **quickcheck** is reminiscent of the ideas explored in *Proofs and Refutations* [13], Lakatos's classic book on the logic of mathematical discovery, but with a 21st century technological twist.

As we illustrated with our discussion of SeCaV and NaDeA, the proof assistant based approach makes it easy to show students more of the breadth of modern logic: here that there may be multiple (very different) proof systems, that these are all interrelated, and that results from one setting can sometimes be usefully applied in another. Using a proof assistant helps bring this abstract fact to life. This is partly because you can cover more ground quickly and yet precisely, and (once again) partly because of the flexibility they offer: logic is being taught not just as something to be learned, but as material that can be moulded, played with — and passed on to others.

But perhaps the most powerful point about using proof assistants to teach logics lies in the most obvious fact of all: to work with a proof assistant is to *do* logic — and indeed, to do *metalogic*. Learning logic this way is like learning a language by immersion — done well, it can be fast and deep. One of the questions we find most interesting is: how far can it be pushed? Our students have typically been computer science and mathematics students, but it is easy to point to topics in linguistics and philosophy that could benefit from being taught with the aid of proof assistants, for example, reference and inference in natural language [4], belief revision [10], and logical dynamics [1]. These are relatively new fields that draw heavily on mathematical and logical ideas. In all of them there are a variety of approaches; sometimes the link between them are well-understood technically, other times less so. What all three areas have in common is that they would benefit from access to flexible mechanisms for logical exploration. The use of proof assistant technology could lift the burden here for a new generation of students, and open the door to new understanding in these areas.



# Appendix: Example Derivations

Full versions of 3 proofs discussed in the paper.

**proposition** ⟨ (∀*x. p x*) ⟶ *p a* ∧ *p b* ⟩ **by** *metis*

**lemma** ⟨ ⊩
[
 *Imp* (*Uni* (*Pre* ''*p*'' [*Var 0*])) (*Con* (*Pre* ''*p*'' [*Fun* ''*a*'' []]) (*Pre* ''*p*'' [*Fun* ''*b*'' []]))
]
⟩
**proof** −
 **from** *AlphaImp* **have** *?thesis* **if** ⟨ ⊩
  [
   *Neg* (*Uni* (*Pre* ''*p*'' [*Var 0*])),
   *Con* (*Pre* ''*p*'' [*Fun* ''*a*'' []]) (*Pre* ''*p*'' [*Fun* ''*b*'' []])
  ]
  ⟩
  **using** *that* **by** *simp*
 **with** *Ext* **have** *?thesis* **if** ⟨ ⊩
  [
   *Con* (*Pre* ''*p*'' [*Fun* ''*a*'' []]) (*Pre* ''*p*'' [*Fun* ''*b*'' []]),
   *Neg* (*Uni* (*Pre* ''*p*'' [*Var 0*]))
  ]
  ⟩
  **using** *that* **by** *simp*
 **with** *BetaCon* **have** *?thesis* **if** ⟨ ⊩
  [
   *Pre* ''*p*'' [*Fun* ''*a*'' []],
   *Neg* (*Uni* (*Pre* ''*p*'' [*Var 0*]))
  ]
  ⟩ **and** ⟨ ⊩
  [
   *Pre* ''*p*'' [*Fun* ''*b*'' []],
   *Neg* (*Uni* (*Pre* ''*p*'' [*Var 0*]))
  ]
  ⟩
  **using** *that* **by** *simp*
 **with** *Ext* **have** *?thesis* **if** ⟨ ⊩
  [
   *Neg* (*Uni* (*Pre* ''*p*'' [*Var 0*])),
   *Pre* ''*p*'' [*Fun* ''*a*'' []]
  ]
  ⟩ **and** ⟨ ⊩
  [
   *Neg* (*Uni* (*Pre* ''*p*'' [*Var 0*])),
   *Pre* ''*p*'' [*Fun* ''*b*'' []]
  ]
  ⟩
  **using** *that* **by** *simp*



```
      with GammaUni have ?thesis if ⟨⊩
        [
          Neg (Pre ''p'' [Fun ''a'' []]),
          Pre ''p'' [Fun ''a'' []]
        ]
        ⟩ and ⟨⊩
        [
          Neg (Pre ''p'' [Fun ''b'' []]),
          Pre ''p'' [Fun ''b'' []]
        ]
        ⟩
        using that by simp
      with Ext have ?thesis if ⟨⊩
        [
          Pre ''p'' [Fun ''a'' []],
          Neg (Pre ''p'' [Fun ''a'' []])
        ]
        ⟩ and ⟨⊩
        [
          Pre ''p'' [Fun ''b'' []],
          Neg (Pre ''p'' [Fun ''b'' []])
        ]
        ⟩
        using that by simp
      with Basic show ?thesis
        by simp
    qed
```

---

**proposition** ⟨(∀x. p x ⟶ q x) ⟶ (∃x. p x) ⟶ (∃x. q x)⟩ **by** *metis*

**lemma** ⟨⊩
  [
    *Imp*
      (*Uni* (*Imp* (*Pre* ''p'' [*Var 0*]) (*Pre* ''q'' [*Var 0*])))
      (*Imp* (*Exi* (*Pre* ''p'' [*Var 0*])) (*Exi* (*Pre* ''q'' [*Var 0*])))
  ]
  ⟩
**proof** −
  **from** *AlphaImp* **have** *?thesis* **if** ⟨⊩
    [
      *Neg* (*Uni* (*Imp* (*Pre* ''p'' [*Var 0*]) (*Pre* ''q'' [*Var 0*]))),
      *Imp* (*Exi* (*Pre* ''p'' [*Var 0*])) (*Exi* (*Pre* ''q'' [*Var 0*]))
    ]
    ⟩
    **using** *that* **by** *simp*
  **with** *Ext* **have** *?thesis* **if** ⟨⊩
    [
      *Imp* (*Exi* (*Pre* ''p'' [*Var 0*])) (*Exi* (*Pre* ''q'' [*Var 0*])),
      *Neg* (*Uni* (*Imp* (*Pre* ''p'' [*Var 0*]) (*Pre* ''q'' [*Var 0*])))
    ]
    ⟩
    **using** *that* **by** *simp*



  **with** *AlphaImp* **have** *?thesis* **if** ⟨⊩
  [
    *Neg* (*Exi* (*Pre* ′′*p*′′ [*Var 0*])),
    *Exi* (*Pre* ′′*q*′′ [*Var 0*]),
    *Neg* (*Uni* (*Imp* (*Pre* ′′*p*′′ [*Var 0*]) (*Pre* ′′*q*′′ [*Var 0*])))
  ]
  ⟩
  **using** *that* **by** *simp*
**with** *DeltaExi* **have** *?thesis* **if** ⟨⊩
  [
    *Neg* (*Pre* ′′*p*′′ [*Fun* ′′*a*′′ []]),
    *Exi* (*Pre* ′′*q*′′ [*Var 0*]),
    *Neg* (*Uni* (*Imp* (*Pre* ′′*p*′′ [*Var 0*]) (*Pre* ′′*q*′′ [*Var 0*])))
  ]
  ⟩
  **using** *that* **by** *simp*
**with** *Ext* **have** *?thesis* **if** ⟨⊩
  [
    *Neg* (*Uni* (*Imp* (*Pre* ′′*p*′′ [*Var 0*]) (*Pre* ′′*q*′′ [*Var 0*]))),
    *Neg* (*Pre* ′′*p*′′ [*Fun* ′′*a*′′ []]),
    *Exi* (*Pre* ′′*q*′′ [*Var 0*])
  ]
  ⟩
  **using** *that* **by** *simp*
**with** *GammaUni* **have** *?thesis* **if** ⟨⊩
  [
    *Neg* (*Imp* (*Pre* ′′*p*′′ [*Fun* ′′*a*′′ []]) (*Pre* ′′*q*′′ [*Fun* ′′*a*′′ []])),
    *Neg* (*Pre* ′′*p*′′ [*Fun* ′′*a*′′ []]),
    *Exi* (*Pre* ′′*q*′′ [*Var 0*])
  ]
  ⟩
  **using** *that* **by** *simp*
**with** *Ext* **have** *?thesis* **if** ⟨⊩
  [
    *Exi* (*Pre* ′′*q*′′ [*Var 0*]),
    *Neg* (*Imp* (*Pre* ′′*p*′′ [*Fun* ′′*a*′′ []]) (*Pre* ′′*q*′′ [*Fun* ′′*a*′′ []])),
    *Neg* (*Pre* ′′*p*′′ [*Fun* ′′*a*′′ []])
  ]
  ⟩
  **using** *that* **by** *simp*
**with** *GammaExi* **have** *?thesis* **if** ⟨⊩
  [
    *Pre* ′′*q*′′ [*Fun* ′′*a*′′ []],
    *Neg* (*Imp* (*Pre* ′′*p*′′ [*Fun* ′′*a*′′ []]) (*Pre* ′′*q*′′ [*Fun* ′′*a*′′ []])),
    *Neg* (*Pre* ′′*p*′′ [*Fun* ′′*a*′′ []])
  ]
  ⟩
  **using** *that* **by** *simp*



  **with** *Ext* **have** *?thesis* **if** ⟨⊩
    [
    *Neg* (*Imp* (*Pre* ''*p*'' [*Fun* ''*a*'' []]) (*Pre* ''*q*'' [*Fun* ''*a*'' []])),
    *Pre* ''*q*'' [*Fun* ''*a*'' []],
    *Neg* (*Pre* ''*p*'' [*Fun* ''*a*'' []])
    ]
    ⟩
    **using** *that* **by** *simp*
  **with** *BetaImp* **have** *?thesis* **if** ⟨⊩
    [
    *Pre* ''*p*'' [*Fun* ''*a*'' []],
    *Pre* ''*q*'' [*Fun* ''*a*'' []],
    *Neg* (*Pre* ''*p*'' [*Fun* ''*a*'' []])
    ]
    ⟩ **and** ⟨⊩
    [
    *Neg* (*Pre* ''*q*'' [*Fun* ''*a*'' []]),
    *Pre* ''*q*'' [*Fun* ''*a*'' []],
    *Neg* (*Pre* ''*p*'' [*Fun* ''*a*'' []])
    ]
    ⟩
    **using** *that* **by** *simp*
  **with** *Ext* **have** *?thesis* **if** ⟨⊩
    [
    *Pre* ''*p*'' [*Fun* ''*a*'' []],
    *Neg* (*Pre* ''*p*'' [*Fun* ''*a*'' []])
    ]
    ⟩ **and** ⟨⊩
    [
    *Pre* ''*q*'' [*Fun* ''*a*'' []],
    *Neg* (*Pre* ''*q*'' [*Fun* ''*a*'' []])
    ]
    ⟩
    **using** *that* **by** *simp*
  **with** *Basic* **show** *?thesis*
    **by** *simp*
**qed**

---

**proposition** ⟨∃*x*. ∀*y*. *p y* ∨ ¬ *p x*⟩ **by** *metis*
**lemma** ⟨⊩
  [
  *Exi* (*Uni* (*Dis* (*Pre* ''*p*'' [*Var 0*]) (*Neg* (*Pre* ''*p*'' [*Var 1*]))))
  ]
  ⟩
**proof** −
  **from** *Ext* **have** *?thesis* **if** ⟨⊩
    [
    *Exi* (*Uni* (*Dis* (*Pre* ''*p*'' [*Var 0*]) (*Neg* (*Pre* ''*p*'' [*Var 1*])))),
    *Exi* (*Uni* (*Dis* (*Pre* ''*p*'' [*Var 0*]) (*Neg* (*Pre* ''*p*'' [*Var 1*]))))
    ]
    ⟩
    **using** *that* **by** *simp*



**with** *GammaExi*[**where** *t*=⟨ *Fun* ″*a*″ [] ⟩] **have** *?thesis* **if** ⟨⊩
  [
  *Uni* (*Dis* (*Pre* ″*p*″ [*Var 0*]) (*Neg* (*Pre* ″*p*″ [*Fun* ″*a*″ []]))),
  *Exi* (*Uni* (*Dis* (*Pre* ″*p*″ [*Var 0*]) (*Neg* (*Pre* ″*p*″ [*Var 1*]))))
  ]
  ⟩
  **using** *that* **by** *simp*
**with** *DeltaUni* **have** *?thesis* **if** ⟨⊩
  [
  *Dis* (*Pre* ″*p*″ [*Fun* ″*b*″ []]) (*Neg* (*Pre* ″*p*″ [*Fun* ″*a*″ []])),
  *Exi* (*Uni* (*Dis* (*Pre* ″*p*″ [*Var 0*]) (*Neg* (*Pre* ″*p*″ [*Var 1*]))))
  ]
  ⟩
  **using** *that* **by** *simp*
**with** *AlphaDis* **have** *?thesis* **if** ⟨⊩
  [
  *Pre* ″*p*″ [*Fun* ″*b*″ []],
  *Neg* (*Pre* ″*p*″ [*Fun* ″*a*″ []]),
  *Exi* (*Uni* (*Dis* (*Pre* ″*p*″ [*Var 0*]) (*Neg* (*Pre* ″*p*″ [*Var 1*]))))
  ]
  ⟩
  **using** *that* **by** *simp*
**with** *Ext* **have** *?thesis* **if** ⟨⊩
  [
  *Exi* (*Uni* (*Dis* (*Pre* ″*p*″ [*Var 0*]) (*Neg* (*Pre* ″*p*″ [*Var 1*])))),
  *Pre* ″*p*″ [*Fun* ″*b*″ []]
  ]
  ⟩
  **using** *that* **by** *simp*
**with** *GammaExi*[**where** *t*=⟨ *Fun* ″*b*″ [] ⟩] **have** *?thesis* **if** ⟨⊩
  [
  *Uni* (*Dis* (*Pre* ″*p*″ [*Var 0*]) (*Neg* (*Pre* ″*p*″ [*Fun* ″*b*″ []]))),
  *Pre* ″*p*″ [*Fun* ″*b*″ []]
  ]
  ⟩
  **using** *that* **by** *simp*
**with** *DeltaUni* **have** *?thesis* **if** ⟨⊩
  [
  *Dis* (*Pre* ″*p*″ [*Fun* ″*c*″ []]) (*Neg* (*Pre* ″*p*″ [*Fun* ″*b*″ []])),
  *Pre* ″*p*″ [*Fun* ″*b*″ []]
  ]
  ⟩
  **using** *that* **by** *simp*
**with** *AlphaDis* **have** *?thesis* **if** ⟨⊩
  [
  *Pre* ″*p*″ [*Fun* ″*c*″ []],
  *Neg* (*Pre* ″*p*″ [*Fun* ″*b*″ []]),
  *Pre* ″*p*″ [*Fun* ″*b*″ []]
  ]
  ⟩
  **using** *that* **by** *simp*



**with** *Ext* **have** *?thesis* **if** ⟨⊩
  [
    *Pre* ″*p*″ [*Fun* ″*b*″ []],
    *Neg* (*Pre* ″*p*″ [*Fun* ″*b*″ []])
  ]
  ⟩
  **using** *that* **by** *simp*
 **with** *Basic* **show** *?thesis*
  **by** *simp*
**qed**

# Acknowledgements

We thank Alexander Birch Jensen and Anders Schlichtkrull for discussions.